\documentclass[aps,prl,preprint,superscriptaddress,showpacs,amssymb]{revtex4}
\usepackage{bm}
\usepackage{epsfig}
\usepackage{amssymb}
\usepackage{bm}

\begin{document}
\title{Oscillation of
spin polarization in a two-dimensional hole gas under a
perpendicular magnetic field}

\author{P.~Kleinert}%
\affiliation{Paul-Drude-Institut f\"ur Festk\"orperelektronik,
Hausvogteiplatz 5-7, 10117 Berlin, Germany}%
\author{V.V.~Bryksin}%
\affiliation{A.F. Ioffe Physical Technical Institute,
Politekhnicheskaya 26, 194021 St. Petersburg, Russia}

\begin{abstract}
Spin-charge coupling is studied for a strongly confined
two-dimensional hole gas subject to a perpendicular magnetic
field. The study is based on spin-charge coupled drift-diffusion
equations derived from quantum-kinetic equations in an exact
manner. The spin-orbit interaction induces an extra out-of-plane
spin polarization. This contribution exhibits a persistent
oscillatory pattern in the strong-coupling regime.
\end{abstract}

\pacs{72.25.Dc, 73.63.Hs, 85.75.-d}

\maketitle

\section{Introduction}
Recently, the study of spin-polarized transport in semiconductors
has received much attention because of its potential applications
in the field of semiconductor spintronics. Many authors have
focused on spin-orbit interaction (SOI) that allows for purely
electric manipulation of spin polarization in semiconductors.
Beside this useful feature, SOI brings also into play the
undesired spin relaxation due to the coupling between the momentum
of charge carriers and their spin (c.f., for instance, Ref.
\cite{Vasil_2004}). Owing to this inhomogeneous broadening, each
elastic and inelastic scattering mechanism opens up a spin
dephasing channel \cite{PRB_2945}. The character of spin
relaxation is quite different in systems with weak and strong SOI
\cite{PRB_195329}. In the latter case, the magnetization can
oscillate even in the absence of external fields. In contrast, for
weakly spin-orbit coupled systems, the spin polarization decays
exponentially unless it is permanently stimulated by external
fields.

The decay of spin polarization seems to be unavoidable because of
the non-conservation of the total spin. Nevertheless, a special
persistent spin-precession pattern has been identified recently
\cite{PRL_236601}. The infinite spin lifetime of this persistent
spin helix occurs in a combined Rashba-Dresselhaus model at a
certain wave vector that gives rise to a special spin rotation
symmetry. Furthermore, oscillations of the nonequilibrium spin
density in real space, which is induced by the Rashba SOI, have
been reported in a number of recent papers
\cite{PRB_033316,PRB_195308,PRB_205307}. These results on robust
spin oscillations certainly encourage further experimental and
theoretical studies of long-lived spin coherence states
\cite{PRB_155317} in semiconductors with SOI.

In this paper, we focus on a strongly confined two-dimensional
hole gas (2DHG) and study the mutual influence of SOI and a
perpendicular external magnetic field. It is well known that a
quantizing perpendicular magnetic field appreciably changes the
transport properties of a two-dimensional electron gas (2DEG). The
quantized energy spectrum manifests in Shubnikov-de Haas
oscillations of the resistivity and may lead to the quantum-Hall
effect. Due to the SOI-induced splitting and crossing of Landau
levels, a beating pattern arises in Shubnikov-de Haas oscillations
\cite{PRB_085313}, which is used to determine the SOI strength
from the measured magnetoresistivity. Similar quantum oscillations
have been identified in the spin-relaxation rate
\cite{PRB_245312}. Other studies
\cite{PRB_085302,PRB_045303,Zhang_L477} deal with the combined
effects of Rashba and Dresselhaus SOI on the magnetotransport in a
2DEG. Unfortunately, comparable investigations of a 2DHG are
limited although the SOI is much stronger in such systems. We
mention the analysis of transport equations for the 2DHG at zero
magnetic field \cite{PRB_193316}, the study of spin dephasing in
p-type semiconductor quantum wells \cite{PRB_125314}, and the
treatment of the spin-Hall effect \cite{PRB_115333}.

Our work is aimed to study  the spin-charge coupled motion of
holes in narrow quantum wells subject to a perpendicular magnetic
field. Based on a rigorous density-matrix approach, spin-charge
coupled drift-diffusion equations are derived for the 2DHG. In
order to focus on general physical properties of the SOI in
semiconductors, we adopt the simple cubic Rashba model that has
been used in the
literature~\cite{Gerchikov,PRB_155303,APL_3151,PRB_085308,PRB_155314}
to simulate the SOI in a 2DHG. This model has the striking
peculiarity that there is no coupling between the spin and charge
components of the density matrix. One should contrast this finding
with the linear Rashba model, which is used to study effects of
SOI in a 2DEG. In this model, the SOI leads to a coupling between
spin and charge degrees of freedom. For a 2DHG such a coupling is
exclusively induced by external fields. Here, we treat a magnetic
field applied perpendicular to the layer. Due to this field, the
charge density and out-of-plane spin polarization couple to each
other in the 2DHG. Consequently, an inhomogeneous spin
polarization induces charge gradients, which are accompanied by an
induced internal electric field calculated via Poisson's equation.
The most interesting feature of our approach is, however, the
observation that the character of the magnetic-field-induced
spin-charge coupling differs qualitatively in the weak and strong
coupling regime. For weak SOI, the dephasing time becomes much
larger than the momentum-relaxation time so that the dominating
mechanism is spin diffusion. In this regime, the field-induced
magnetization exhibits only a smooth exponential dependence on
spatial coordinates. Conversely, for strong SOI, the ballistic
spin-transport regime is established, in which oscillations of the
out-of-plane magnetization can occur. An experimental verification
of this prediction would facilitate the technological exploitation
of these long-lived spin states for the fabrication of logical
gates.

\section{Basic theory}
We treat coupled spin-charge excitations on the basis of an
effective-mass Hamiltonian, which refers to the heavy-hole band of
thin p-type quantum wells and which has been adopted in the
literature~\cite{Gerchikov,PRB_155303,APL_3151,PRB_085308,PRB_155314}
as an acceptable simple approximation. Our model includes
short-range spin-independent elastic scattering on impurities and
a constant perpendicular magnetic field $B$, from which only the
Zeeman splitting is considered. The related heavy-hole Hamiltonian
of the cubic Rashba model has the second-quantized form
\begin{equation}
H=\sum_{\bm{k},\lambda }a_{\bm{k}\lambda }^{\dag}\left[ \varepsilon_{%
\bm{k}}-\varepsilon _{F}\right] a_{%
\bm{k}\lambda }-\sum_{\bm{k},\lambda ,\lambda ^{\prime }}\left(
\hbar\vec{\bm{\omega}}_{
\bm{k}} \cdot \vec{\bm{\sigma }}_{\lambda \lambda ^{\prime }}\right) a_{\bm{k}%
\lambda }^{\dag}a_{\bm{k}\lambda ^{\prime }}+
u\sum\limits_{{\bm{k}},{\bm{k}}^{\prime}}
\sum\limits_{\lambda}a_{{\bm{k}}\lambda}^{\dag}a_{{\bm{k}}^{\prime}\lambda},
\label{Hamil}
\end{equation}
where $a_{\bm{k}\lambda}^{\dag}$ ($a_{\bm{k}\lambda}$) denote the
creation (annihilation) operators with in-plane quasi-momentum
$\bm{k}=(k_x,k_y,0)$ and spin $\lambda$. In Eq.~(\ref{Hamil}), we
introduced the Fermi energy $\varepsilon_F$, the vector of Pauli
matrices $\vec{\bm{\sigma}}$, and the strength $u$ of the
'white-noise' elastic impurity scattering, which gives rise to the
momentum relaxation time $\tau$. The heavy-hole band is described
by the dispersion relation
$\varepsilon_{\bm{k}}=\hbar^2{\bm{k}}^2/(2m)$. The coupling of
spin states as described by
\begin{equation}
\hbar\vec{\bm{\omega}}_{\bm{k}}=
\left[i\frac{\alpha}{2}(k_{+}^3-k_{-}^3),%
\frac{\alpha}{2}(k_{+}^3+k_{-}^3),\hbar\omega_c \right],
\label{omega}
\end{equation}
is due to the Zeeman splitting  $\hbar\omega_c=g^*\mu_BB/2$ and
the SOI, the strength of which is denoted by $\alpha$. In
Eq.~(\ref{omega}), we have $k_{\pm}=k_{x}\pm i k_{y}$,
$k_x=k\cos(\varphi)$, $k_y=k\sin(\varphi)$, and
$\hbar\omega_{k}=\alpha k^3$. Within the Born approximation with
respect to elastic impurity scattering, the four components
$(f,\vec{\bm{f}})=(\sum_{\lambda}f_{\lambda}^{\lambda},\sum_{\lambda,\lambda'}f_{\lambda'}^{\lambda}
\vec{\sigma}_{\lambda\lambda'})$ of the spin-density matrix
$f_{\lambda'}^{\lambda}$ satisfy the following Laplace-transformed
quantum-kinetic equations \cite{PRB_165313,PRB_205317}
\begin{equation}
sf-\frac{i\hbar}{m}(\bm{\kappa}\cdot\bm{k})f+i\vec{\bm{\omega}}_{\bm{\kappa}}
({\bm{k}})\cdot\vec{\bm{f}} =\frac{1}{\tau}(\overline{f}-f)+f_0,
\label{kin1}
\end{equation}
\begin{eqnarray}
&&s\vec{\bm{f}}+2(\vec{\bm{\omega}}_{\bm{k}}\times\vec{\bm{f}})
-\frac{i\hbar}{m}(\bm{\kappa}\cdot\bm{k})\vec{\bm{f}}
+i\vec{\bm{\omega}}_{\bm{\kappa}} ({\bm{k}})f\nonumber\\
&&=\frac{1}{\tau}(\overline{\vec{\bm{f}}}-\vec{\bm{f}})+\frac{1}{\tau}
\frac{\partial}{\partial\varepsilon_{\bm{k}}}
\overline{f\hbar\vec{\bm{\omega}}_{\bm{k}}}-
\frac{\hbar\vec{\bm{\omega}}_{\bm{k}}}{\tau}
\frac{\partial}{\partial\varepsilon_{\bm{k}}} \overline{f}
+\vec{\bm{f}}_0, \label{kin2}
\end{eqnarray}
in which the SOI-dependent vector
\begin{equation}
\hbar\vec{\bm{\omega}}_{\bm{\kappa}}({\bm{k}})=3\alpha
\left[(k_y^2-k_x^2)\kappa_y-2k_xk_y\kappa_x,(k_x^2-k_y^2)\kappa_x-2k_xk_y\kappa_y,0
\right]
\end{equation}
couples the spin and charge degrees of freedom to each other. The
wave vector ${\bm{\kappa}}$ refers to the center-of-mass motion
and disappears in models that refer to homogeneous spin and charge
distributions. Initial charge and spin densities are denoted by
$f_0=n$ and $\vec{\bm{f}}_0$, respectively. The cross line over
$\bm{k}$-dependent functions indicates an integration over the
polar angle $\varphi$ of the in-plane vector $\bm{k}$. $s$ denotes
the variable of the Laplace transformation and takes over the role
of the time parameter $t$.

By treating the kinetic Eqs.~(\ref{kin1}) and (\ref{kin2}) in the
long-wavelength limit, coupled spin-charge drift-diffusion
equations are derived for the angle-averaged spin-density matrix
$(\overline{f},\overline{\vec{\bm{f}}})$. The method has already
been applied to a 2DEG without any external
fields~\cite{PRB_205317}. In this approach, it is assumed that
carriers quickly reestablish thermal equilibrium. This fact
justifies the ansatz
$\overline{f}(\varepsilon_{\bm{k}},{\bm{\kappa}}\mid
s)=n(\varepsilon_{\bm{k}})F({\bm{\kappa}}\mid s)$, where
$n(\varepsilon_{\bm{k}})$ denotes the Fermi distribution function.
Expanding the solution of Eqs.~(\ref{kin1}) and (\ref{kin2}) up to
second order in ${\bm{\kappa}}$ and calculating the integral over
the angle $\varphi$, we obtain our main theoretical result namely
the following spin-charge coupled drift-diffusion equations
\begin{equation}
(s+D_0\kappa^2)\overline{f}-\Gamma_z\kappa^2\overline{f}_z=n,
\label{e1}
\end{equation}
\begin{equation}
(s+\frac{1}{\tau_{sz}}+D_z\kappa^2)\overline{f}_z+\Gamma_0
\overline{f}=f_{z0}, \label{e2}
\end{equation}
\begin{equation}
(\sigma_0^2s\tau+2\Omega^2(2s\tau+1))\overline{f}_{x}
+D_{x}\tau\kappa^2\overline{f}_{x}-2\sigma_0\omega_c\tau(1+\widetilde{D}\tau\kappa^2)\overline{f}_y
=(\sigma_0^2+2\Omega^2)\tau f_{x,0}, \label{e3}
\end{equation}
\begin{equation}
(\sigma_0^2s\tau+2\Omega^2(2s\tau+1))\overline{f}_{y}
+D_{x}\tau\kappa^2\overline{f}_{y}+2\sigma_0\omega_c\tau(1+\widetilde{D}\tau\kappa^2)\overline{f}_x
=(\sigma_0^2+2\Omega^2)\tau f_{y,0}, \label{e4}
\end{equation}
where we used the abbreviations $\sigma_0=s\tau+1$ and
$\Omega=\omega_k\tau$. The $k$-dependent coefficients in this set
of equations have the form
\begin{equation}
D_0=\frac{D}{\sigma_0^2},\quad
\Gamma_z=24\frac{\hbar\omega_c\tau}{m}\Omega^2
\frac{\sigma_0^2+2\Omega^2}{\sigma_0^2(\sigma_0^2+4\Omega^2)^2},
\end{equation}
\begin{equation}
\frac{1}{\tau_{sz}}=\frac{4\Omega^2}{\sigma_0\tau},\quad
D_z=D\frac{\sigma_0^2-12\Omega^2}{(\sigma_0^2+4\Omega^2)^2}, \quad
\Gamma_0=-\frac{\chi H}{\sigma_0\mu_B\tau_{sz}}, \label{Dz}
\end{equation}
\begin{equation}
D_x=D\frac{\sigma_0^6+24\sigma_0^2\Omega^4+32\Omega^6}{\sigma_0^2(\sigma_0^2+4\Omega^2)^2},\quad
\widetilde{D}=D\frac{4\Omega^2-3\sigma_0^2}{(\sigma_0^2+4\Omega^2)^2},
\end{equation}
where we introduced the diffusion coefficient $D=v^2\tau/2$, the
Bohr magneton $\mu_B$, and magnetic susceptibility $\chi$. The
Eqs.~(\ref{e1}) to (\ref{e4}) completely decouple in the absence
of the external magnetic field, when
$\Gamma_z=\Gamma_0=\omega_c=0$. This is a peculiarity of the cubic
Rashba model. When a perpendicular magnetic field is applied to
the 2DHG, a steady-state out-of-plane spin polarization arises
\begin{equation}
f_z^{(0)}=-\hbar\omega_cn^{\prime}=\frac{\chi H}{\mu_B},
\end{equation}
which couples to the charge density. For a 2DEG the situation is
different. In this case, the out-of-plane spin polarization
couples to the in-plane spin components~\cite{PRB_205317}. The
most surprising feature of our solution exhibits the
spin-diffusion coefficient $D_z$ in Eq.~(\ref{Dz}), the form of
which agrees with a recently published result
\cite{PRB_193316,PRB_205317} derived by an alternative approach.
This particular diffusion coefficient becomes negative for strong
SOI ($\Omega>\sigma_0/\sqrt{12}$) indicating an instability of the
spin system. In this regime, spin diffusion has the tendency to
strengthen initial spin fluctuations. The competition between this
self-strengthening and spin relaxation leads to undamped spin
oscillations that are characteristic for ballistic spin transport.
Such spin oscillations result from the coupling between the charge
density and the out-of-plane spin polarization expressed by
Eqs.~(\ref{e1}) and (\ref{e2}). What is interesting is that this
unusual result for $D_z$ can only be obtained by taking into
account the off-diagonal elements of the density matrix. (In fact,
neglecting $f_x$ and $f_y$ in Eq.~(\ref{kin2}), we obtain simply
$D_z=D$). Therefore, the oscillations in the strong SOI regime
have a pure quantum-mechanical origin that is manifested in the
quasi-classical Eqs.~(\ref{e1}) and (\ref{e2}). Strictly speaking,
this result arises beyond the applicability of the drift-diffusion
approach~\cite{PRB_205317,PRB_125307}.

The time dependence of the in-plane spin polarization as described
by Eqs.~(\ref{e3}) and (\ref{e4}) is governed by characteristic
poles \cite{PRL_226602,PRB_165313} that are calculated from
$\sigma_0^2s\tau+2\Omega^2(2s\tau+1)=0$. Let us treat the
strong-coupling regime $\Omega\gg 1$ for the in-plane spin
polarization that is determined by poles at $s\tau=-3/4\pm
2i\Omega$. Performing the inverse Laplace and Fourier
transformations, we obtain for the spectral spin polarization the
result
\begin{equation}
\overline{f}_x({\bm{r}},k\mid
t)=\exp\left[-\frac{\bm{r}^2}{16Dt}-\frac{3t}{4\tau}
\right]\biggl\{\frac{\cos(2\omega_kt)}{t/\tau}f_{x0}-\frac{\omega_c\tau}{2\Omega}
\sin(2\omega_kt)f_{y0} \biggl\}/(32\pi D),
\end{equation}
\begin{equation}
\overline{f}_y({\bm{r}},k\mid
t)=\exp\left[-\frac{\bm{r}^2}{16Dt}-\frac{3t}{4\tau}
\right]\biggl\{\frac{\cos(2\omega_kt)}{t/\tau}f_{y0}+\frac{\omega_c\tau}{2\Omega}
\sin(2\omega_kt)f_{x0} \biggl\}/(32\pi D),
\end{equation}
which describes damped oscillations of an initially at
${\bm{r}}={\bm{0}}$ injected spin packet. The external magnetic
field couples initial nonvanishing in-plane spin components to
each other. A spot like initial in-plane spin polarization could
be produced in experiment by a short laser pule. The evolution of
this initial inhomogeneous spin distribution is described by
Eqs.~(14) and (15).

\section{Spin polarization for a stripe geometry}
In this Section, the magnetic-field-induced coupling between the
charge distribution $\overline{f}$ and the out-of-plane spin
polarization $\overline{f}_z$  in a 2DHG is treated in more detail
for a stripe of width $2L$ oriented along the $x$ axis. To this
end, the steady-state solution ($s=0$, $\sigma_0=1$) of
Eqs.~(\ref{e1}) and (\ref{e2}) is transformed back to the
representation in spatial coordinates $x$ and $y$. Due to the
considered stripe geometry, the densities are independent of $x$.
The variation of the charge density $\overline{f}(k,y)$ induces a
self-consistent internal electric field $E_y(k,y)$ that is
calculated from the Poisson equation. This internal in-plane
electric field is a by-product of the spin-charge coupling. Its
reaction on the spin is accounted for by drift terms in
Eqs.~(\ref{e1}) and (\ref{e2}). Its phenomenological consideration
in Eq.~(\ref{e1}) for the carrier density is ruled by the concept
of effective chemical potential \cite{PRB_165313}. Motivated by
studies of electric-field effects on spin transport, we introduce
a similar contribution in Eq.~(\ref{e2}) for the out-of-plane spin
polarization. Putting all together, the following set of coupled
equations for spin-charge excitations are obtained
\begin{equation}
D(k)\overline{f}^{\prime}(k,y)-\mu
E_y(k,y)\overline{f}(k,y)-\Gamma_z(k)\overline{f}_z^{\prime}(k,y)=0,
\end{equation}
\begin{equation}
D_z(k)\overline{f}_z^{\prime\prime}(k,y)-\mu
E_y(k,y)\overline{f}_z^{\prime}(k,y)
-\frac{1}{\tau_{sz}(k)}\overline{f}_z(k,y)-\Gamma_0(k)\overline{f}(k,y)=0,
\end{equation}
\begin{equation}
E_y^{\prime}(k,y)=\frac{4\pi
e}{\varepsilon}(\overline{f}(k,y)-n(k)),
\end{equation}
in which $\mu=e\tau/m$ denotes the mobility and $\varepsilon$ is
the dielectric constant. Primes indicate derivatives with respect
to $y$. We derive an analytical solution of these equations by
calculating the lowest-order contributions in the induced electric
field $E_y$. Within this perturbational schema, we make the ansatz
$\overline{f}=n+\Delta \overline{f}$ and
$\overline{f}_z=f_z^{(0)}+\Delta \overline{f}_z$, where the
corrections result from the spin-charge coupling $\Delta
\overline{f}$, $\Delta \overline{f}_z\sim E_y$. In addition,
hard-wall boundary conditions $\Delta \overline{f}_z(\pm L)=0$ and
the existence of interface charges $E_y(\pm L)=\pm E_0$ are
assumed. We obtain the analytic solution
\begin{equation}
{\Delta \overline{f}_z}=\frac{\lambda_1\lambda_2E_0\Gamma_0}{4\pi
e/\varepsilon}\frac{\cosh(\lambda_2L)\cosh(\lambda_1y)-\cosh(\lambda_1L)\cosh(\lambda_2y)}{N(L)}
,\, \frac{E_y}{E_0}=\frac{N(y)}{N(L)},\label{e19}
\end{equation}
where $\lambda_{1,2}$ are calculated from the secular equation
\begin{equation}
\left(D\lambda^2 -\frac{4\pi e}{\varepsilon}\mu
n\right)\left(D_z\lambda^2-\frac{1}{\tau_{sz}}\right)-\lambda^2\Gamma_0\Gamma_z=0.
\end{equation}
In Eq.~(\ref{e19}), the abbreviation
\begin{equation}
N(y)=\lambda_2\left(D_z\lambda_1^2-\frac{1}{\tau_{sz}}\right)\cosh(\lambda_2L)\sinh(\lambda_1y)
-\lambda_1\left(D_z\lambda_2^2-\frac{1}{\tau_{sz}}\right)\cosh(\lambda_1L)\sinh(\lambda_2y)
\end{equation}
was introduced. For weak magnetic fields $\omega_c\tau\ll 1$, we
obtain the final result
\begin{equation}
\Delta f_z(\varepsilon_{\bm{k}},y)=-\frac{eE_0\chi
H\lambda_1^2L_D^3}{\mu_B(1-(\lambda_1L_D)^2)}
\coth(L/L_D)\biggl\{\frac{\cosh(y/L_D)}{\cosh(L/L_D)}
-\frac{\cosh(\lambda_1y)}{\cosh(\lambda_1L)} \biggl\}
\frac{dn(\varepsilon_{\bm{k}})}{d\varepsilon_{\bm{k}}}
\label{final},
\end{equation}
with
\begin{equation}
\lambda_1=1/\sqrt{D_z\tau_{sz}},\quad \lambda_2=\sqrt{4\pi e\mu
n/(D\varepsilon)}=L_D^{-1},
\end{equation}
where $L_D$ denotes the Debye screening length. Again, we meet a
peculiarity of the cubic Rashba model for a 2DHG. The final
integral over the energy $\varepsilon_{\bm{k}}$ is easily
calculated at low temperatures. Due to the factor
${dn(\varepsilon_{\bm{k}})}/{d\varepsilon_{\bm{k}}}$, the
field-induced spin polarization is exclusively determined by
energies at the Fermi surface for a degenerate hole gas.
Therefore, the recently studied inhomogeneous broadening
\cite{PRB_125314} due to elastic scattering is ineffective in this
regime. However, inelastic scattering, which we disregarded in
this work, may play an essential role for the formation of a
persistent oscillatory spin pattern at strong SOI.

The character of the solution for the out-of-plane spin
polarization mainly depends on the strength of the SOI.
%%%%%%%%%%%%%%%%%%%%%%%%%%%%%%%
\begin{figure}
%%%%%\vspace*{5cm}
\centerline{\epsfig{file=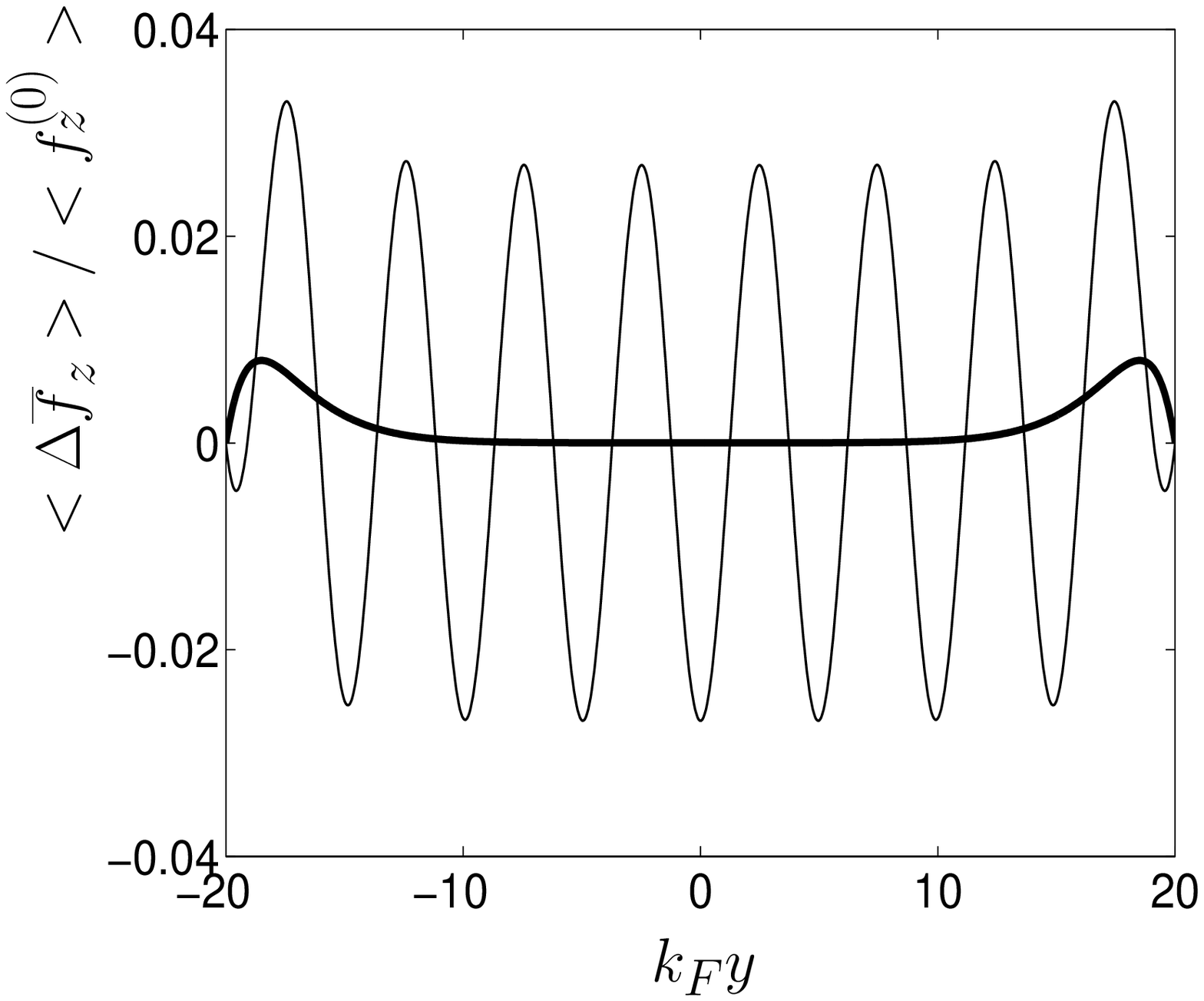,width=8.5cm}}
\caption{SOI-induced out-of-plane spin polarization obtained from
Eq.~(\ref{final}) by integrating over $\varepsilon_{\bm{k}}$
(indicated by $<\dots >$). At zero temperature, all quantities are
calculated at the Fermi momentum $k_F$. Parameters used in the
calculation are: $k_F=0.1$~nm$^{-1}$, $m\alpha/\hbar^2=2$~nm,
$E_0=100$~V/cm, and $n=10^{15}$/cm$^3$. The thick and thin lines
refer to weak ($\Omega=0.173$, $\tau=0.05$~ps) and strong
($\Omega=0.34$, $\tau=0.1$~ps) spin-orbit coupling, respectively.}
\label{fig1}
\end{figure}
% %%%%%%%%%%%%%%%%%%%%%%%%%%%%%
In weakly coupled systems ($\Omega<1/\sqrt{12}$, $D_z>0$), the
spin polarization exhibits an exponential dependence as shown by
the thick line in Fig.~1. The self-consistent coupling between
spin and charge degrees of freedom leads to an excess
magnetization at the boundaries of the stripe. The picture changes
dramatically, when we consider the strong-coupling regime
($\Omega>1/\sqrt{12}$, $D_z<0$). In this case, the wave number
$\lambda_1$ becomes imaginary giving rise to spin-coherent
oscillations. An example for this persistent spin pattern is shown
by the thin line in Fig.~1. Despite the included elastic
scattering on impurities, the spin lifetime of these oscillations
is infinite in the strong-coupling regime. Moreover, the
oscillation amplitude is considerably enhanced at the resonance
$\lambda_1L=(2n+1)\pi/2$ with $n$ being any integer. A similar
enhancement has been predicted for the SOI-induced zitterbewegung
\cite{PRL_206801}. This observation also remind of a Fabry-Perot
interferometer in optics. Like the finesse of the interferometer
diverges for perfect reflective mirrors, the amplitude of the spin
oscillations becomes infinite for the above mentioned particular
values of the spin-orbit coupling and the width of the stripe.
This idealized behavior indicates that beside elastic scattering
on impurities also other spin-relaxation mechanisms have to be
taken into account for a more realistic description of spin
excitations at strong SOI. The experimental observation of the
interesting persistent oscillatory spin structure is certainly
challenging. It requires a spin detection set up with a high
spatial resolution (the typical wavelength of the oscillations is
of the order of $100$~nm). As the magnetic field leads to a
coupling between the out-of-plane spin polarization and the charge
density, both the induced internal electric field and the charge
density exhibit similar oscillations in the strong coupling
regime.

\section{Summary}
We studied a 2DHG with SOI and elastic impurity scattering under
the influence of a perpendicular magnetic field. Applying an exact
procedure, spin-charge coupled drift-diffusion equations were
derived from quantum-kinetic equations for the spin-density
matrix. The magnetic field mainly causes a coupling between the
out-of-plane spin polarization and the charge density. The
character of effects that result from this coupling strongly
depend on the strength of the SOI. For weak SOI
($\Omega<1/\sqrt{12}$), spin diffusion gives rise to an
exponential decay of an initial spin polarization. In contrast,
for strong SOI, the spin transport exhibits ballistic character so
that oscillations of the magnetization can occur. This general
conclusion was illustrated by a treatment of the spin polarization
in a stripe composed of a 2DHG. The magnetic field induces a
background magnetization that is superimposed by a contribution
stemming from the SOI. The excess magnetization, which results
from the spin-charge coupling, exhibits a persistent oscillatory
spin pattern for systems with strong spin-orbit coupling. Similar
standing and propagating spin oscillations with wavelength down to
several nanometers have been treated for thin magnetic film
samples \cite{Kruglyak}. The application of this mechanism for
spin-wave logic gates depends on whether short-wavelength spin
oscillations can be manipulated and detected by a suitable
experimental set up.
\begin{acknowledgments}
This work was supported by the Deutsche Forschungsgemeinschaft and
the Russian Foundation of Basic Research.%
\end{acknowledgments}

%\newpage
\section*{References}
% BibTeX users please use
%\bibliographystyle{prsty}
%\bibliography{abbrev,spin}

\end{document}